\documentclass[preprint,11pt]{aastex}

\renewcommand{\deg}{\ifmmode^\circ\else$^\circ$\fi}
\newcommand{\et}{{\it et al.}~}
\newcommand{\vs}{{\it vs.}}

\newcommand{\arcm}{\hbox{$^\prime$}}
\newcommand{\arcs}{\hbox{$^{\prime\prime}$}}

\newcommand{\um}{\hbox{$\mu$m}}
\newcommand{\xx}{$\times$}
\newcommand{\nWmmsr}{\hbox{nW m$^{-2}$ sr$^{-1}$}}
\newcommand{\MJysr}{\hbox{MJy sr$^{-1}$}}
\newcommand{\kJysr}{\hbox{kJy sr$^{-1}$}}

\newcommand{\KkJyval}{16.4} \newcommand {\KkJysig}{4.4}
\newcommand{\LkJyval}{12.8} \newcommand {\LkJysig}{3.8}
\newcommand{\LkJy}{\hbox{$\LkJyval \pm \LkJysig$~\kJysr}}
\newcommand{\KkJy}{\hbox{$\KkJyval \pm \KkJysig$~\kJysr}}
\newcommand{\LnW}{\hbox{$11.0 \pm 3.3$~\nWmmsr}}
\newcommand{\KnW}{\hbox{$22.4 \pm 6.0$~\nWmmsr}}

\begin{document}

\title{Tentative Detection of the Cosmic Infrared Background at 2.2 
and 3.5 \um\ Using Ground Based and Space Based Observations}

\author{V.  Gorjian} \affil{Jet Propulsion Laboratory, California Institute of
Technology, 4800 Oak Grove Dr., Pasadena, CA  91109}  \email{vg@jpl.nasa.gov}

\and

\author{E.  L.  Wright and R.  R.  Chary} \affil{Department of Physics and
Astronomy, University of California, Los Angeles, CA  90095-1562}
\email{wright@astro.ucla.edu, rchary@astro.ucla.edu}

\begin{abstract}

The Cosmic InfraRed Background (CIRB) is the sum total of
the redshifted and reprocessed short wavelength radiation from the era of
galaxy formation, and hence contains vital information about the history of
galactic evolution.
One of the main problems 
associated with estimating an isotropic CIRB in
the near infrared (1-5 \um) is the unknown
contribution from stars within our own galaxy.
The optimal observational window to search for 
a background in the near-IR is at 3.5 \um\ since that is
the wavelength region where the  other main foreground, the zodiacal dust 
emission, is the least. 
It is not possible to map out the entire 3.5 \um\ sky
at a resolution which will accurately estimate the flux from stars.
However, since the CIRB is presumably isotropic, it can potentially be
detected by selecting a smaller field and imaging it at good resolution
to estimate the stellar intensity.
We selected a 2\deg\xx2\deg~ ``dark spot" near the North Galactic Pole 
which had the least intensity at 3.5 \um\ after a zodiacal light
model was subtracted from the all-sky maps generated by the Diffuse InfraRed
Background Experiment (DIRBE).
Still, the large area of the field made it very difficult to mosaic 
at 3.5  \um\ using the
available arrays. Thus, the field was mosaiced at 2.2 \um, then the bright
stars  were selected and re-imaged at 2.2 and 3.5 \um. The resulting total 
intensity of the bright  stars was combined with a model for the contribution 
from dimmer stars and subtracted from the  zodi-subtracted DIRBE map. 
The contribution from the interstellar medium was also subtracted
leaving a residual intensity at 2.2 \um\ of:
\KkJy\ or \KnW, and at 3.5 \um\ of:
\LkJy\ or \LnW.
The nature of our analysis suggests that this  excess emission is
probably a detection of the cosmic background in the near infrared.

\end{abstract}

\keywords{cosmology:  observations --- diffuse radiation --- infrared:general}

\section{Introduction}

One of the few tools available for studying the ``Dark Ages'' following the
era of matter-radiation decoupling is the Cosmic InfraRed Background
(CIRB) radiation.  The CIRB is the result of the cumulative, short
wavelength emissions from pregalactic, protogalactic, and galactic systems
which through dust reprocessing and cosmological redshifting are now at
infrared wavelengths. 
Although the importance of looking for an extragalactic infrared background
has been discussed for some time \citep{PP67,LT68,Pe69,Ha70,Ka76}, 
relatively little theoretical attention had been paid to it due to the
difficulties inherent in trying to observe and verify theoretical
predictions.  But advances in infrared technology have stimulated an
increasing amount of interest in trying to determine the characteristics of
the CIRB \citep{BCH86,McD86,Fab87,Fab88,BCH91,Lo95, MS98, Pei99}.
The theoretical work indicates that measuring the spectral intensity and the
anisotropy of the CIRB will have important implications regarding the amount
of matter undergoing luminous episodes in the pregalactic Universe, the
nature and evolution of those luminosity sources, the nature and
distribution of cosmic dust, and the density and luminosity evolution of
infrared-bright galaxies (for review see \citet{Ha96}).

Although technology is no longer a hindrance in observing the CIRB, there
are other difficulties in trying to see this cosmic relic.  The bright
foreground from the atmosphere of the earth, the dust in the solar system,
and the stellar and interstellar emissions of our own galaxy combine to make
detecting the CIRB a formidable task.

Several experiments have been carried out to try to determine the CIRB from
within the atmosphere. \citet{Ma88} and \citet{No92} used 
rocket borne cameras to try to escape the contamination from the
atmosphere and observe the CIRB.  They had only limited success and could set
upper limits on the CIRB without any detections.  Lower limits were determined
from deep galaxy counts conducted in the near-IR.  These two limits began
setting the first tight constraints on galaxy evolution models.

The logical next step in trying to detect the CIRB was to go into space to
try and eliminate the more local foreground; thus, the Cosmic Background
Explorer (COBE) satellite carried with it an experiment specifically
designed to detect the CIRB \citep{Bo92}.  The Diffuse InfraRed Background 
Experiment
(DIRBE) was cryogenically cooled, thus allowing it to observe multiple
infrared wavelengths without much instrumental background. Having eliminated
the effect of the earth's atmosphere and the instrumental background, only
the zodiacal dust in the solar system and the stellar and interstellar
emissions of the Galaxy remained.

\begin{figure}[ptb]
\plotone{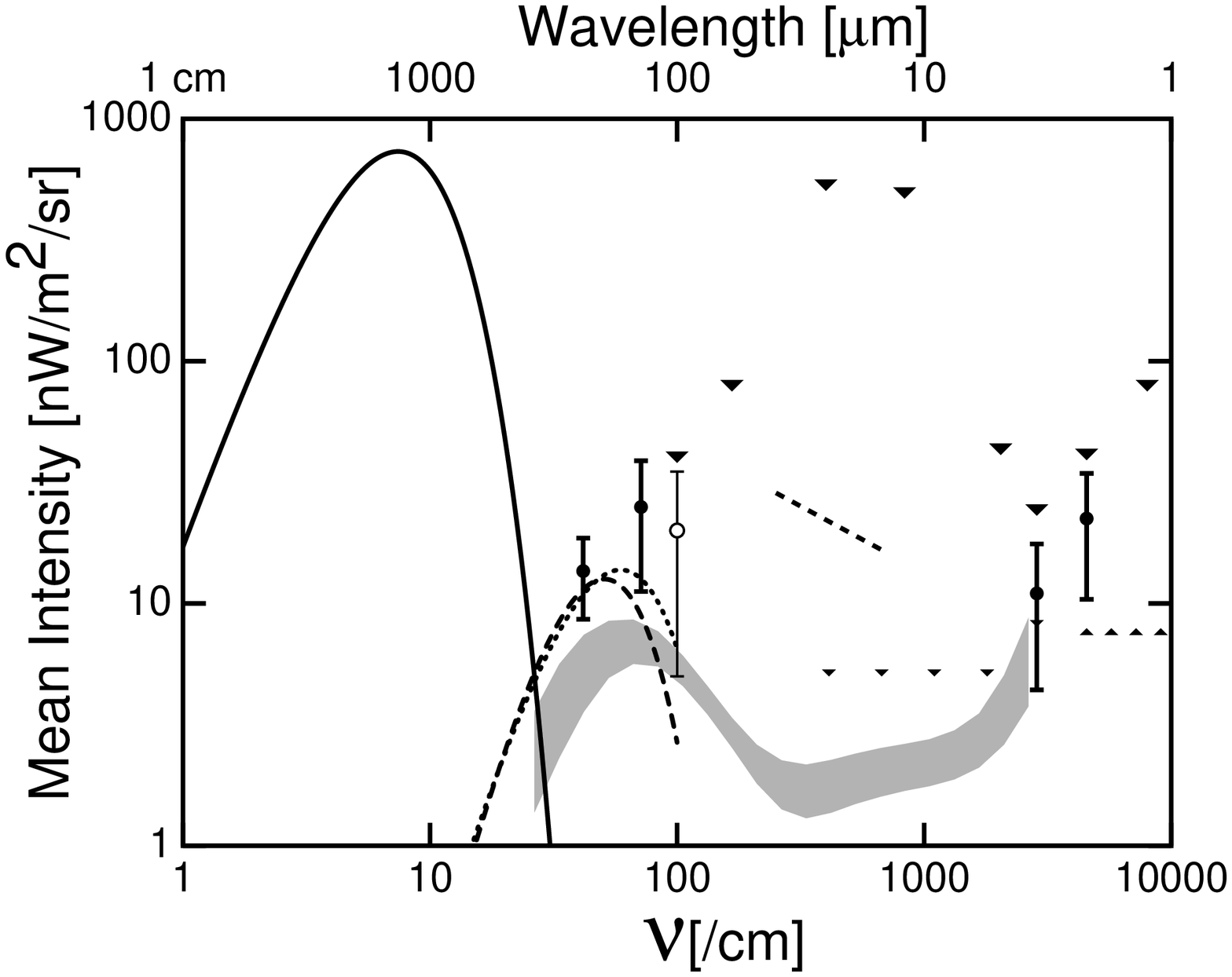}
\caption{
Limits and detections by DIRBE compared to various CIRB models.
The DIRBE 2$\sigma$ upper limits at 1.25-60 \um\ are represented
by solid downward triangles. The detections at
2.2 \um\ and 3.5 \um\ (this paper), 
and at 140 \um\ and 240 \um\ \citep{HAKDO98} are represented by
solid circles with $\pm$2$\sigma$ error bars. 
The 100 \um\ intensity is represented by both the 95\% 
confidence limit (open circle)
on the non-galactic, non-zodiacal intensity of 
5-35 \nWmmsr\ and the quoted upper limit on the cosmic background.
The Far InfraRed Absolute
Spectrophotometer (FIRAS) 125-5000 \um\ detection \protect\citep{FDMBS98} 
is shown by a dotted curve while the \protect\citet{LABDP99} is shown
by a dashed curve; the $\lambda < 2.2\;\mu$m  
lower limits of \protect\citet{PMZFB98} 
are represented by small upward triangles. 
The upper limits on the CIRB based on TeV
observations of Mrk 501 
\protect\citep{SF98} 
are shown as small downward triangles,
while the ``detection'' based on TeV observations of Mrk 421 
\citep{DS94} 
is shown as a dashed line.
The gray band shows the models of
\citet{MS98}. 
\label{fig:dwek}}
\end{figure}

Even with the contamination of the atmosphere removed, the DIRBE data did
not readily reveal an isotropic background.  Removing the remaining 
foregrounds turned out
to be a difficult task with the zodiacal dust proving to be especially
difficult as none of the models were able to completely eliminate it from
the data,  especially in the near and mid-infrared where the dust emits and
scatters the most.  The emissions from interstellar dust were more readily
modeled and removed and the emissions from stars were also modeled and
removed.  These attempts at foreground removal finally yielded values for
a CIRB at 140 \um\ and 240 \um, first as a possible detection 
from an independent group 
\citep{SFD98} 
and then the definitive value from the DIRBE
science team 
\citep{HAKDO98}. 
These values are shown in Figure \ref{fig:dwek} in relation to various
models.  As can be seen, the models are within factors of 2-6 of the
observed values at 140 \um\ and 240 \um, indicating a convergence between
observation and theory in the far-IR.

Yet in the near-IR (NIR) (1 - 5 \um), DIRBE was unable to detect the CIRB. 
The contamination from the zodiacal dust as well as uncertainties in the 
contribution from galactic stars resulted in residuals
which had large error bars and thus could only be considered as upper 
limits \citep{HAKDO98}. 
In addition, the residual maps in the NIR failed the tests for isotropy 
even in limited
regions of the sky. 

\subsection{Zodiacal Light Removal}

The zodiacal dust is made up of dust from asteroids and comets and extends
out to the asteroid belt.  The particles scatter and emit light from the 
UV through the infrared with a minimum of the sum of scattering and
emission at 3.5 \um. In recent decades a great deal has been added to our
knowledge of the zodiacal dust, especially with the advent of satellites
like IRAS and COBE. Taking the sum of our knowledge today, we know that the 
zodiacal dust is distributed in a fan shape centered on the sun and 
that it is slightly tilted with  respect to the ecliptic.  
There are dust bands associated with asteroidal
families, and there is a zodiacal dust ring at a distance of 1.01 AU. 
Unfortunately this accumulated knowledge has not brought with it a 
completely accurate model of the
dust distribution.  The DIRBE science team has produced a model of the dust
distribution 
\citep{KWFRA98} 
to subtract from the DIRBE all-sky maps, but the model parameters are not
unique and thus, after subtraction, the model zodiacal light leaves some
residual effects.

The \citet{KWFRA98} model leaves a large residual intensity in the galactic
polar caps at 25 \um, the DIRBE band that is most dominated by the zodiacal
light.  For example, the 25 \um\ intensity toward the DIRBE dark spot at
$(l,b) = (120.8^\circ,65.9^\circ)$ in the DIRBE 
Zodi-Subtracted Mission Average (ZSMA) maps is 1.76 \MJysr.
This cannot be a cosmic background because the lack of $\gamma$-ray emission
toward Mkn 501 limits the CIRB to be $< 33$ \kJysr\
(Funk \et\ 1998).
It also cannot be galactic cirrus because the 100 \um\ intensity in this
field is 1.27 \MJysr\ in the ZSMA maps, and Arendt \et\ (1998) specify
the ISM intensity as $R(\lambda)(I(100)-I_\circ)$, with $R(25) = 0.0480$ and
$I_\circ = 0.66$~\MJysr, so the ISM intensity is 29~\kJysr.  By elimination,
most of this intensity must be zodiacal.

In order to reduce the residual zodiacal emission in the maps, Wright (1997)
added one ``observation'' that the high $b$ intensity at 25 \um\ should be zero
to the more than $10^5$ observations used in the zodiacal model fitting.  
Even this
very low weight pseudo-observation lowered the 25 \um\ intensity in the dark
spot to 0.26 \MJysr.  This indicates that the isotropic component of the
zodiacal emission is very poorly constrained in fits that just look at the
time variation to measure the zodiacal light.  
The Appendix in Wright (1998) discusses zodiacal
light models in more depth, and the Appendix of this paper gives the 
actual parameters of the model we have used.

Since the residual 25 \um\ intensity is now only 1\% of the total zodiacal
emission, we might hope for errors in the ZL model equal to 1\% of the ecliptic
pole intensity.  But the situation is more uncertain at 2.2 and 3.5 \um\ due
to the scattered component of the ZL.  Adjusting the thermal emission component
to fit the 25 \um\ intensity will not necessarily lead to a correct scattered
component.  So we have adopted ZL modeling errors of 5\% of the intensity at
the ecliptic poles at 2.2 and 3.5 \um.  These errors are slightly lower at 2.2
\um\ 
(5.2 {\it vs.} 6 \nWmmsr) and higher at 3.5 \um\ 
(2.8 {\it vs.} 2 \nWmmsr) than the errors adopted by Kelsall \et\ (1998).

\subsection{Interstellar Medium Contribution}

Another foreground is the emission from dust in the interstellar medium (ISM).
This contribution is significant in the FIR but is relatively small in the NIR.
However, it can not be completely neglected. 
The distribution of the dust was modeled by \citet{AOWSH98} by correlating 
the DIRBE 100 \um\ map with external data sets of gas and dust like HI, HII, 
and CO.  \citet{AOWSH98} specify that the ISM subtracted map at wavelength
$\lambda$ is $I_\nu(\lambda,l,b) - R(\lambda)(I_\nu(100,l,b) - I_\circ)$.
$I_\circ = 0.66\;\MJysr$ is the non-zodiacal, non-galactic component of 
the 100 \um\ map.  The coefficient $R(2.2\;\um)$ is zero,
but $R(3.5\;\um) = 0.00183$ \citep{AOWSH98} due to single photon induced
thermal fluorescence in the 3.3 \um\ PAH feature in very small grains.
We have made this correction on a pixel by pixel basis, and it should remove
the effect of galactic cirrus if the global coefficient applies to our spot.
The ISM correction listed in Table \ref{tab:decomp} is the correction 
averaged over the 17 pixels in the ``dark spot''.

\subsection{Foreground Starlight Removal}

Since there is no all-sky map at a sufficiently high resolution in the NIR,
it is difficult to calculate the cumulative intensity of stars at varying
galactic locations.  This calculation is vital in eliminating one of the
main  foregrounds to the CIRB between 1 \um\ and 5 \um.

To get an estimate of the intensity from stars, theoretical models of the 
Galaxy have been constructed from which number counts in any direction can be
derived.  The most recent and successful model of the Galaxy was an
enhancement of the
\citet{WCVWS92} 
model made by 
\citet{Co94} 
who divided the Galaxy into 5 main components:  the disk, the spiral arms,
the molecular ring, the central bulge, and the extended halo, as well as
several minor components:  Gould's belt, local molecular clouds, reflection
nebulae, etc.  In his model, each major component has 82 different stellar
types distributed among the components with weighting techniques designed to
reproduce observations of each component in the NIR. Additionally the
spectral classes are characterized by absolute magnitudes at IR wavelengths,
a magnitude dispersion, a scale height above the galactic plane, and a solar
neighborhood density.   Finally, dust in the Galaxy is modeled using the 
\citet{RL85} 
uniform extinction law.

For the purpose of detecting the CIRB, the model has to be accurate at high
galactic latitudes where the number counts are lowest and present the least
amount of contamination.  This is where the difficulties with the model
become evident.  Since the sun is situated about 15 pc above the disk and we
are looking through the disk {\it and} the halo when we are looking at high
galactic latitudes, any error in the halo-to-disk weighting will generate
some error in  the star counts.  An additional source of error will be the
actual distance of the sun above the disk.  The greater the height above the
disk, the lower the contribution of disk stars when looking North vs. 
South.  These problems affect the counts and the surface brightness
calculations at about the 10\% level 
\citet{Co96}. 

A different source of error arises when the inherently statistical
predictions of the model are subtracted from observed data.  Any model will
not exactly recreate the number counts in the galactic region of interest. 
There will be a variance from the actual number of stars.  This variance is
of little consequence for dim stars since each star contributes relatively
little to the overall intensity, but the variance becomes very important for 
the brighter stars where the flux of each individual star contributes
significantly to the overall intensity.

While the zodiacal light model is intrinsically on the same flux scale as the
DIRBE data,
the relative calibration of the external stellar data for bright stars 
and models for faint stars {\it vs.} the DIRBE data needs to be known with good
accuracy.  The faint star model of \citet{AOWSH98} used fluxes at
zero magnitude of $F_\circ(K) = 612.3\;\mbox{Jy}$ and
$F_\circ(L) = 285\;\mbox{Jy}$ \citep{DIRBE-ExpSup}.
We have checked these values by
extracting the total flux in 16 pixel areas around the bright
stars $\beta$ And, $\alpha$ Tau, $\alpha$ Aur, $\alpha$ Ori,
$\alpha$ Boo, $\alpha$ Her \& $\beta$ Peg.  The median
$F_\circ(K)$ derived from these stars is $F_\circ(K) = 614\;\mbox{Jy}$
with a mean and standard deviation of the mean $616.6 \pm 12.5\;\mbox{Jy}$.
This calibration agrees with \citet{AOWSH98} to better than 1\%.
But at 3.5 \um, the median $F_\circ(L) = 263\;\mbox{Jy}$ and the mean
is $264.9 \pm 3.4\;\mbox{Jy}$.  This is an 8\% discrepancy.  We will use
our median value,  $F_\circ(L) = 263\;\mbox{Jy}$, and specify the changes
created if the \citet{AOWSH98} value is used instead.

Until recently, only this modeling approach was used to eliminate the
galactic  foreground. However, Dwek \& Arendt (1998; DA98) attempted to
address the uncertainty in the number counts of stars using an innovative
technique.  They used the zodiacal light subtracted DIRBE
all-sky maps at 2.2  \um\ to help determine the background at 3.5 \um. 
First they subtracted a background from the 2.2 \um\ data based on deep K
band (2.2 \um) galaxy counts. This then should have left only the stellar
contribution in the 2.2 \um\ maps.   Then the 1.25, 3.5, and 4.9 \um\ maps
were used to determine the proper colors of  the stars which were then
subtracted from the 3.5 \um\ map.  Based on this approach they made a  
tentative claim of detection for the CIRB at 3.5 \um\ of:
\begin{equation}
\nu I_\nu(3.5\;\mu\mbox{m})  =
                   9.9 + 0.312(\nu I_\nu(2.2\;\mu\mbox{m}) - 7.4) \pm 2.9 
\end{equation}
where $\nu I_\nu$ is in units of \nWmmsr. The factor of
0.312 is the  color correction which was found from the correlation of the
2.2~\um\ map with the 3.5~\um\ map, and the $(\nu I_\nu(2.2\;\mu\mbox{m}) - 
7.4)$ is the real value of the CIRB at K minus the assumed value of the 
CIRB of 7.4~\nWmmsr~from the deep galaxy counts.

\subsection{Motivation for this Project}

As mentioned earlier, the strongest constraints available on the CIRB in the 
NIR are the upper limits by the DIRBE science team \citep{HAKDO98}, 
and the tentative detection at 3.5 \um\ by DA98 also based on DIRBE data.
The DIRBE team did not make a detection claim because they did not have a 
significant residual signal
above the uncertainties associated with the stellar subtraction and also 
because they
had not achieved isotropy in their residual maps \citep{AOWSH98}. 
Their stellar subtraction was achieved in 2 steps. 
In the first step they removed all stars brighter than 15 Jy since they 
were bright enough to be identifiable in the low resolution DIRBE image. 
This translated into a brightness cutoff at K of 4$^{th}$ magnitude and at L 
of 3$^{rd}$ magnitude. In the second step they modeled the contribution of 
stars dimmer than the cutoff value using Cohen's model of the Galaxy. 
Our intention was to reduce those modeling uncertainties by imaging one 
section of sky at high resolution to a limit at least 2 orders of magnitude
fainter than the \citet{AOWSH98} bright star cutoff.
This translated into 9$^{th}$ magnitude at K and 8$^{th}$ 
magnitude at L. 
At these limits, the standard deviation in the intensity contributed by 
the Poisson fluctuations in the numbers of faint stars is reduced to a level 16
times smaller than the standard deviation for the \citet{AOWSH98} threshold.

DA98 claimed a tentative detection at 3.5 \um. 
Although they seemed to have isotropy, they did have a low-level large 
scale gradient, probably due to zodiacal light, that remained in their images. 
In addition, their value at 3.5 \um\ was still based on an assumed value of 
the CIRB at 2.2 \um\ obtained from $K-$band galaxy counts. 

Our intention here was to use a much more stringent zodiacal light 
subtraction routine as well as to avoid any assumption about the CIRB at 
other wavelengths.
With these two new approaches in mind we hoped to place a much tighter 
constraint on the CIRB at 2.2 and 3.5 \um.

\section{Observations}

\subsection{The DIRBE Data}

\begin{figure}[tb]
\plotone{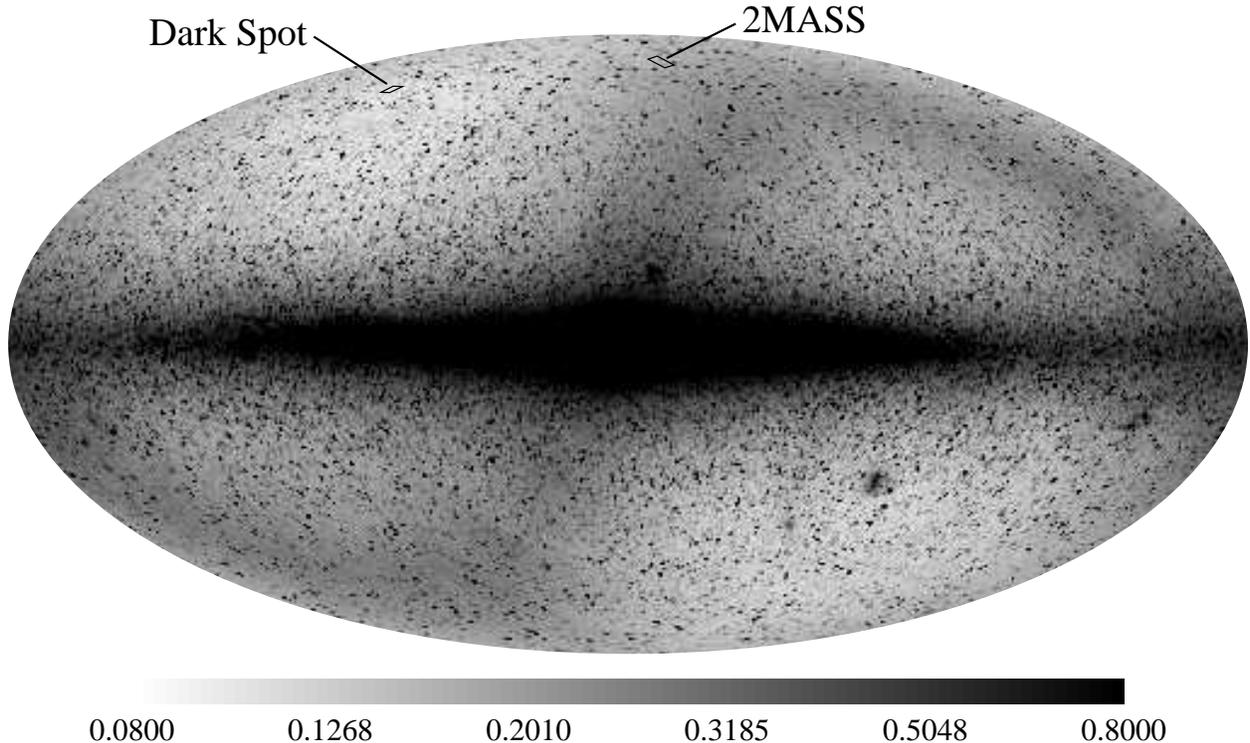}
\caption{
All-sky map from DIRBE at 3.5 
\um\ in galactic coordinates with intensities in \MJysr.
The S-shape of the zodiacal light is evident
as well as the contamination by individual stars which appear 
almost as noise on 
the images. The ``dark spot" ($\alpha = 12^h$ 56.9$^m$, 
$\delta = 51\deg~11\arcm$~(J2000)) and the region used for the 2MASS analysis
($\alpha = 13^h$ 24$^m$, $\delta = 15\deg$~(J2000)) are outlined.
\label{fig:dirbe}}
\end{figure}

The primary dataset for this project was the all-sky map produced by the
DIRBE instrument on the COBE satellite.  DIRBE was helium cooled and
observed the sky at 10 different IR wavebands from 1.25 \um\ to 240 \um. 
The sky was mapped onto 393,216 pixels
with a pixel scale of 0.32\deg~pixel$^{-1}$ and each pixel was observed
several hundred times.  The instrument was also equipped with a chopper
which chopped between the sky and an internal calibrator 32 times per
second.  Finally, the instrument was re-calibrated 5 times per orbit.  All
this redundancy resulted in the best absolutely calibrated map of the
infrared sky obtained to date (Figure \ref{fig:dirbe}).

The first step in determining the CIRB from the DIRBE data was to determine
the location in the sky with the lowest intensity. But since the Earth sits
within the zodiacal dust cloud, the region of lowest emission was still heavily
contaminated by scattering and emission from that dust.  Figure
\ref{fig:dirbe} shows the prominent S-shape (in galactic coordinates) of the
zodiacal dust, and before anything else could be attempted with the data,
that contamination needed to be removed.  To remove the contribution of the
zodiacal light required modeling the distribution of the dust as well as its
emission and scattering characteristics.  This task was accomplished by
\citet{Wr98} resulting in maps where the remaining contamination was from
the Galaxy. We did not use the zodi-subtracted maps  generated by the DIRBE
science team 
\citep{KWFRA98} 
because this model leaves a large residual 25 $\mu$m intensity 
in the region of our ground-based survey.
Thus we used a model
constructed using the ``very strong no-zodi'' condition defined by
\citet{Wr97}, which reduces the residual 25 $\mu$m intensity by a factor
of 7.

From the cleaned map, the area of lowest emission, a ``dark spot", was
chosen  which was a region spanning 2 degrees by 2 degrees.  This
2\deg\xx2\deg~area of  lowest emission turned out to be near the North
Galactic Pole at 
$\alpha = 12^h$ 56.9$^m$, and 
$\delta = 51\deg~11\arcm$~(J2000)\ ($l = 120.8^\circ,\;\;b = 65.9^\circ$).
The low value at this location  represents a combination of the CIRB and the
unresolved stars within the large  (0.7\deg) DIRBE beam.
The next step was then to determine the 
contribution from unresolved stars using higher resolution ground based 
imaging.

\subsection{The Initial Map}

Obtaining the fluxes of stars in the determined ``dark spot'' at 3.5 \um\
was  the primary goal of this project, but mosaicing an area of 4 square
degrees is not  possible  with present day technology. Modern arrays do not
read out fast enough to be  able to image a large region of sky before the
background at 3.5 \um\ saturates the array; therefore, most cameras that
are meant to image in the thermal infrared ($>$ 3 \um) are designed to
either image a very small region of sky to  minimize the background or to
read-out only a small subarray of the detector  chip. 

The main camera available for this project was the UCLA Twin Channel Infrared 
Camera (known locally as Gemini) 
\citep{McL93} 
which mounts onto the Lick 3-m  telescope. On the 3-m, Gemini's two 256\xx256
arrays have identical 3\arcm\xx3\arcm\ fields of view, one array covering the
wavelength range  from 1 to 2.5 \um\ with a  HgCdTe array, and the other
array covering the wavelength range from 1 to 5  \um\ with an InSb array.
The camera employs the subarray technique for readout  when imaging at
wavelengths longer than 3 \um, outputting 64\xx64 pixels giving  it a field
of view of only 45\arcs\xx45\arcs. With a field of view this small  it would
be impossible to mosaic the required area of the ``dark spot" which covers
the equivalent area of 20 full moons.

\begin{figure}[ptb]
\plotone{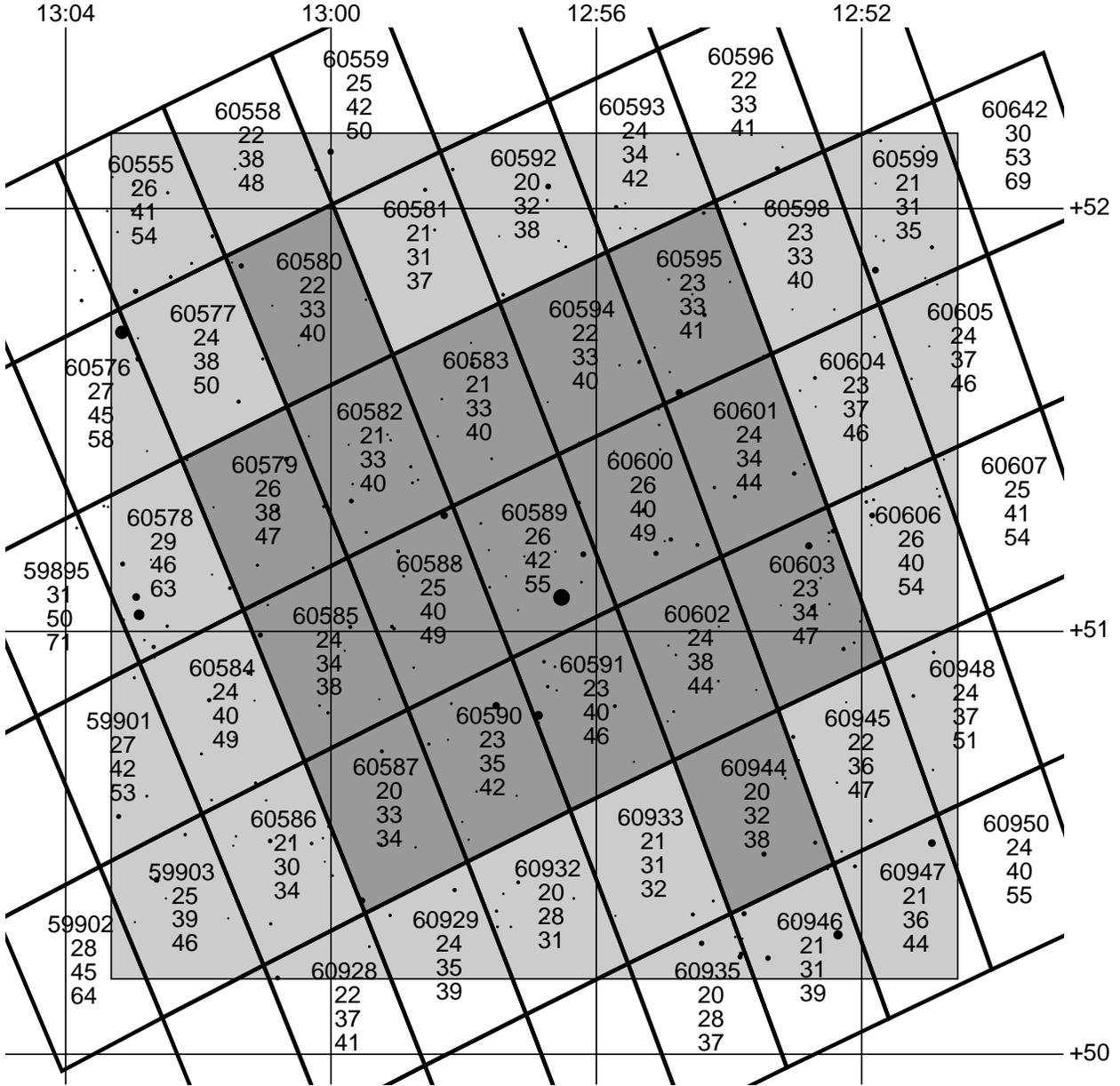}
\caption{
A superposition of the stars within the surveyed
2\deg\xx2\deg~area (thin-lined box gray box) and the DIRBE pixels 
(thick-lined skew rectangles). 
The 17 DIRBE pixels used in our analysis are shaded in  darker gray.
The numbers in 
the pixels are from the top: The DIRBE pixel number and the J, K, and L 
intensities after zodi subtraction
in \kJysr. The grid lines are at even degrees of $\alpha$ and $\delta$. 
The circles are the stars imaged with LIRC2.
The bigger the size, the higher the flux.
\label{fig:cold-spot-list}}
\end{figure}

The solution that was employed was to image the area at K with a different 
camera which due to the lower sky background at K, could have a much larger
field  of  view, then choose the brightest stars from that sample to
re-image at 3.5 \um\ with Gemini on the 3-m. The camera available for this
approach was the Lick  InfraRed Camera 2 (LIRC2). LIRC2 could mount onto the
1-m telescope at Lick and  have a field of view of 7.29\arcm\xx7.29\arcm,
easily allowing the 4 square  degrees to be mapped in a short period of
time. This was done over the nights  of 1997, March 28 and 29. 
The 302 stars catalogued in the LIRC2 survey are shown in Figure
\ref{fig:cold-spot-list}.

The process was complicated during the reduction of the LIRC2  data when it
was discovered that LIRC2 had photometric problems which precluded  highly
accurate photometry. The accuracy was still sufficient to estimate  relative
brightnesses, which was the main result that was required of the data. 
Having obtained the list of stars from the brightest to the dimmest in the 
``dark spot'' we could now proceed with imaging them in the L band.

\subsection{The Main NIR data}

We had three clear and photometric nights at the Lick 3-m
with  Gemini on the nights of 1999, March 26 to 28. Using Gemini's twin
arrays we  imaged our list of stars in both the K band and the L band
simultaneously.

The data were reduced using standard data reduction routines with the 
IRAF\footnote{IRAF ( Image Reduction and Analysis Facility) is distributed
by  the National Optical Astronomy Observatories, which are operated by the 
Association for Research in Astronomy, Inc., under cooperative agreement
with  the National Science Foundation.} software package. Since the sky
brightness  varies dramatically over very short periods of time at 3.5 \um,
differencing of  sequential images was used in the data reduction to remove
detector signature  and sky background. Each star was imaged at 4 different
locations on the array,  while the 10 brightest stars were imaged a total of
12 times.

We also obtained time at the 1.5-m telescope at Palomar on the night of
1999,  April 4, using its IR camera. We imaged the 20 brightest stars at K
band as a  check of our Gemini data as well as imaging the list at J band.
These data were  reduced in a similar fashion to the Gemini data.
The uncertainties for both data sets were determined by the standard 
deviation of the brightness of the standard star throughout the night. 

A total of 77 stars out of the original sample of 302 were observed either 
at Palomar or with Gemini or both.  We have restricted our analysis
to the brightest part of the sample in order to have a reliable
catalog with a well defined limit.
Table \ref{tab:stars} shows the 28 bright stars with $m < 9$
in the dark spot, with the original
LIRC2 K magnitude, the Palomar J and K magnitudes (J60 and K60) and the
Lick Gemini K and L magnitudes.  We have used the Lick K and L magnitudes in
this paper.

\section{Combining the two Datasets}

Having obtained the photometry from the ground for the bright stars, the
task  was to match up the ground based data to the space based data.  Figure
\ref{fig:cold-spot-list} shows a  projection of the DIRBE pixels on the
``dark spot" as well as the stars imaged  by LIRC2.

The most obvious difficulty turns out to be that the DIRBE pixels are not 
aligned with the EW and NS axes; therefore, some only partially enter the 
2\deg\xx2\deg~box. Also the values within the pixels themselves represent 
the average intensity seen when the $0.7^\circ$ square DIRBE beam is
centered within the pixel boundary, and hence are affected by the pixels
around them since the beam does not  abruptly stop at the edge of the pixel.

To compute the response in a given pixel to a given star, we found the 
probability that the star was inside the DIRBE $0.7^\circ$ square beam 
assuming that the beam center was uniformly distributed inside the pixel, 
and that the beam orientation was uniform in angle.  This is a fair
assumption  since each pixel was observed hundreds of times in many
different angles  during the lifetime of the mission.  To be specific, let
$H(\hat{b},\theta,\hat{s})$ be 1 if the star with position given by the unit
vector $\hat{s}$ is in the square beam when it is centered at position
$\hat{b}$ and oriented with position angle $\theta$, and 0 if the star is
not in the beam.  Then the probability of the $j^{th}$ star being in data
taken  in the $i^{th}$ pixel is given by
\begin{equation}
p_{ij} = \frac
{\int_0^{2\pi} \int_{\hat{b} \in \Omega_i} H(\hat{b},\theta,\hat{s}_j) 
	d^2\hat{b} \; d\theta}
	{2\pi \int_{\hat{b} \in \Omega_i} d^2\hat{b}}
\end{equation}
where $\Omega_i$ is the angular extent of the $i^{th}$ pixel.
Then the predicted contribution from bright stars to the intensity in the
$i^{th}$ pixel is given by
\begin{equation}
B_i = \Omega^{-1} \sum_j p_{ij} F_j
\end{equation}
where $\Omega$ is the solid angle of the DIRBE beam and $F_j$ is the flux 
of the $j^{th}$ star. We also need to know
the variance of the intensity in the pixel due to the random orientations
and  random placements of observations within the pixel boundaries.  This can
be obtained from the expected value of $H^2$ within a pixel, but since $H^2
= H$ we find that the variance of $p_{ij}$ is given by $p_{ij}(1-p_{ij})$.
Thus stars at the edge of the beam ($p \simeq 0.5$) make the greatest
contribution to the variance, a phenomenon known as ``flicker noise''. We
then get
\begin{equation}
\sigma(B_i)^2 = \Omega^{-2} \sum_j [p_{ij}(1-p_{ij}) + 
p_{ij}^2\epsilon^2] F_j^2
\end{equation}
where $\epsilon$ is an uncertainty in cross-calibrating the DIRBE and
ground-based data for individual stars
which also includes an allowance for variability in the
stars.   We use $\epsilon^2 = 0.1$ in this work. 
Many IR bright stars are long period variables and will not have the same
flux during the DIRBE observations as they have nine years later during our
ground-based survey, so a fairly large value of $\epsilon^2$ is appropriate.
Note that the variance sum
is dominated by the brightest stars, and is minimized in the ``dark spot''.

\begin{figure}[ptb]
\plotone{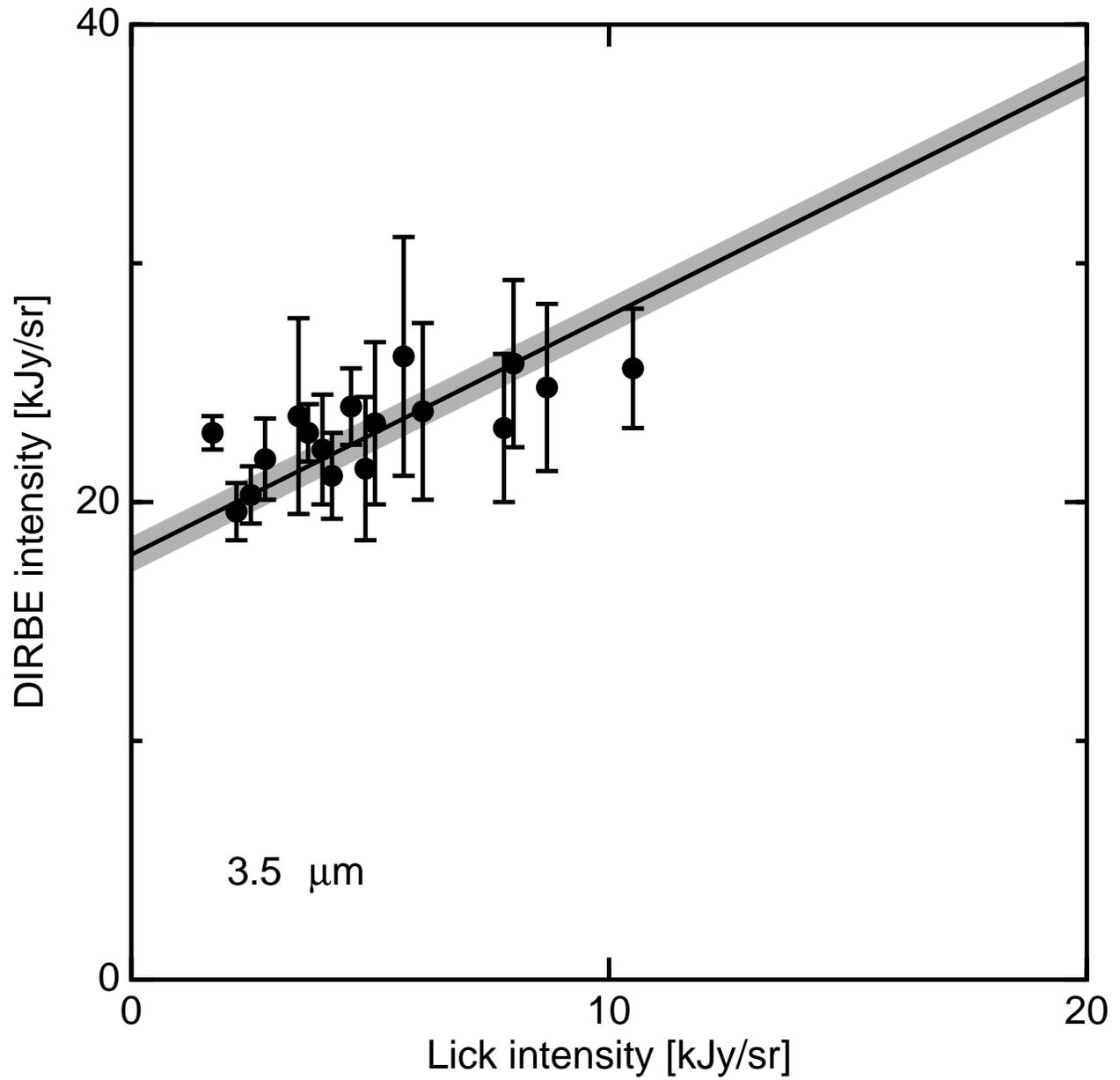}
\caption{
The zodi-subtracted DIRBE 3.5 \um\ 
intensity plotted {\it vs.} the predicted intensity from bright stars based 
on Lick data for stars with $L < 9$ mag 
in the dark spot at $(l,b) = (120.8^\circ,65.9^\circ)$. 
The line is of unit slope.
\label{fig:LSQ-cold-spot-L}}
\end{figure}

\begin{figure}[ptb]
\plotone{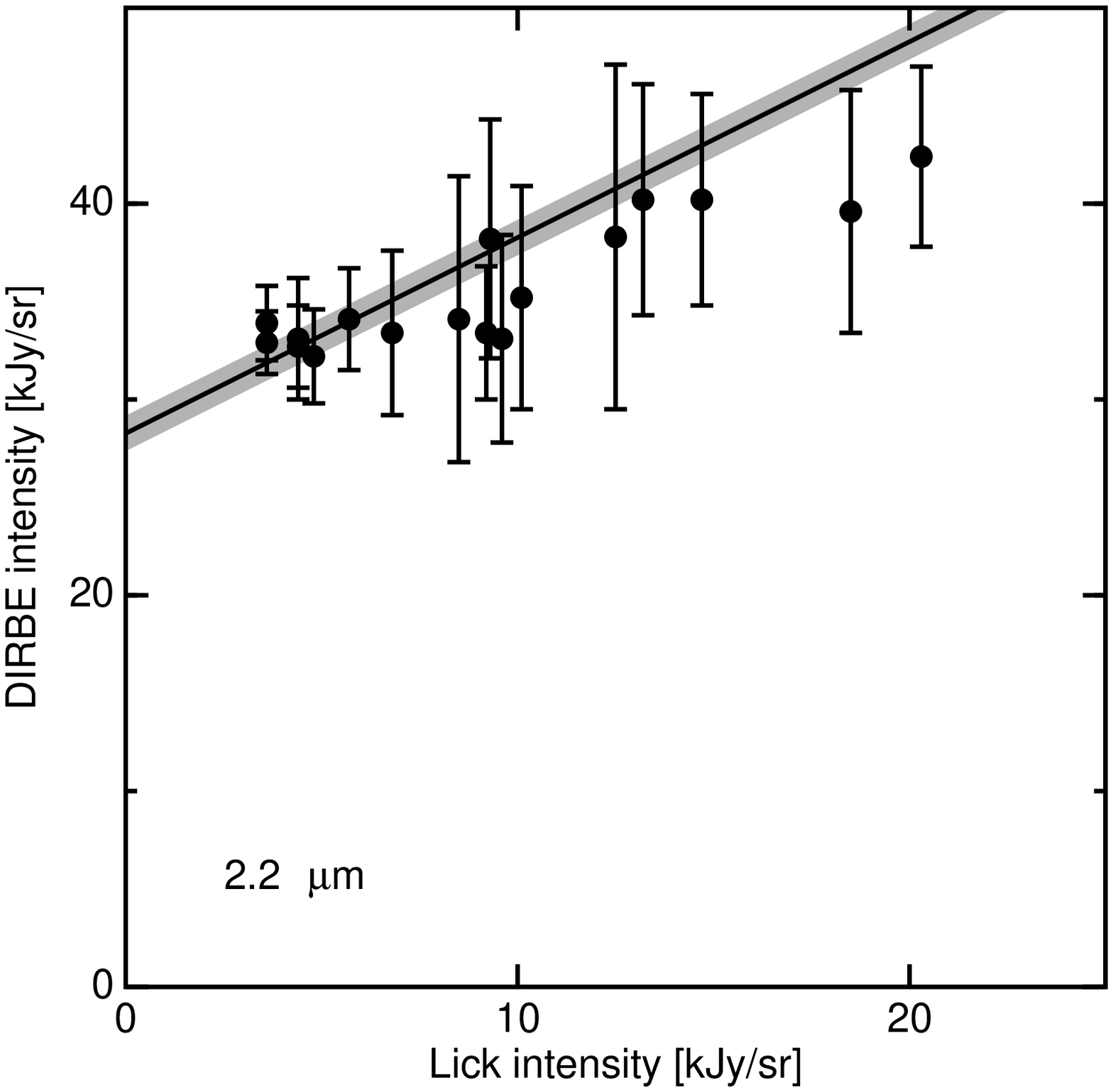}
\caption{
The zodi-subtracted DIRBE 2.2 \um\ 
intensity plotted {\it vs.} the predicted intensity from bright stars based 
on Lick data for stars with $K < 9$ mag
in the dark spot at $(l,b) = (120.8^\circ,65.9^\circ)$. 
The line is of unit slope.
\label{fig:LSQ-cold-spot-K}}
\end{figure}

\begin{figure}[ptb]
\plotone{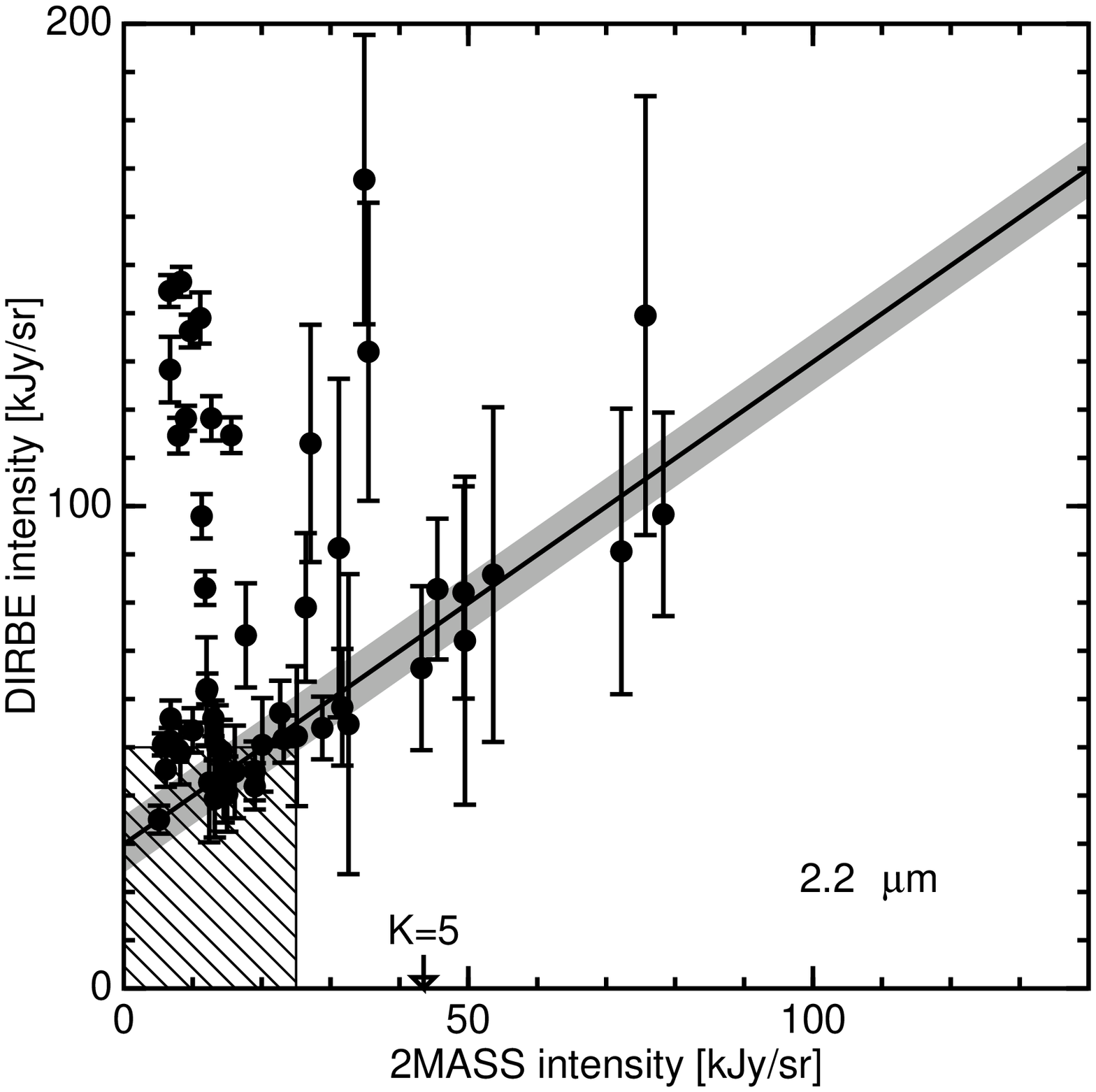}
\caption{
The zodi-subtracted DIRBE 2.2 \um\
intensity plotted {\it vs.} the predicted intensity from bright stars based 
on 2MASS stars with $K < 9$ mag
in the region at $(l,b) = (337.1^\circ,75.6^\circ)$.
The arrow labeled K=5 shows the intensity 
produced by a single $5^{th}$ magnitude star in the DIRBE beam.
The cross-hatched area in the lower left shows the plotting box for
Figure \protect\ref{fig:LSQ-cold-spot-K}.
The line is of unit slope and the intercept is in good agreement with 
our value for the intensity from stars $K > 9$ mag plus the CIRB.
\label{fig:LSQ-201d-15d-K}}
\end{figure}

We have evaluated the bright star contribution for all stars with $K < 9$
mag or $L < 9$ mag in the dark spot. Note that the cutoff magnitude does not 
come from LIRC2 so as to avoid any additional error being introduced because 
of LIRC2's photometric inaccuracy in the final result. 
Stars were detected below these limits with accurate photometry but were not 
used, with the conservative cut-off being at 9th magnitude. 
There are only 17 DIRBE pixels whose
response to a uniform illumination pattern is at least 90\% due to radiation
from within the dark spot, and these 17 pixels form our DIRBE sample. Given
the DIRBE data $D_i$ and the bright star intensity $B_i$, we can estimate
the cosmic infrared background using $C_i = D_i-B_i-F-Z$, where $F$ is the
contribution from faint stars evaluated from star count models, and $Z$ is
the zodiacal contribution.  Both $F$ and $Z$ are essentially constant over
our spot and thus are not subscripted by pixel.  We combine the several
estimates of the cosmic infrared background using a weighted median:  we
find the value $\hat{C}$ that minimizes
\begin{equation}
E = \sum_i \left\vert\frac{C_i-\hat{C}}{\sigma(B_i)}\right|
\end{equation}
The uncertainty in $\hat{C}$ is found using the half-sample method
\citep{BF96}: 
many Monte Carlo runs were performed each with a randomly chosen half of the
pixels, and the scatter in these determinations gives the uncertainty in
$\hat{C}$. The result for the CIRB plus the faint stars ($\hat{C}+F$) at 
K is $28.27 \pm 0.90\;\kJysr$,
and at L is $17.81 \pm 0.75 \; \kJysr$.
These uncertainties are based only on the statistical pixel to pixel scatter 
within the ``dark spot'', but they do include the fluctuations in the faint
star contribution. Figure \ref{fig:LSQ-cold-spot-L} shows the
zodi-subtracted DIRBE 3.5 $\mu$m intensity, $D_i-Z$, plotted against the
bright star intensity predicted from the Lick L band data. The slope of a
line fitted to these points is only 0.4, but this does not mean that the
DIRBE calibration differs by a factor of 0.4 from the ground-based
calibration.  Rather it shows that the variance of the stellar contribution 
is not dominated by the bright stars in this ``dark spot'', and thus the
predicted stellar intensity is a noisy estimator of the actual stellar
intensity. In this circumstance, the slope of a least-squares line is 
biased low, as shown in Figure 2 of \citet{Wr92}. 
Figure \ref{fig:LSQ-cold-spot-K} shows the zodi-subtracted 
DIRBE 2.2 \um\ intensity \vs\ the bright star intensity predicted from the 
Lick K band data.
Figure \ref{fig:LSQ-201d-15d-K} shows a similar plot using 2MASS data in a
$2.4^\circ \times 4^\circ$ region centered at $\alpha = 13^h\;24^m$, $\delta
= 15^\circ$. This region is at high galactic latitude 
($l = 337^\circ,\;\;b=76^\circ$) but has not been
selected to have very few bright stars.
Even though pixels affected by stars listed as saturated in the
2MASS catalog have been deleted, there appear to be
several stars that are partially saturated in
the 2MASS data leading to pixels with high DIRBE intensities but low 2MASS
intensities. 
However, the plotted line with unit slope is clearly consistent with
a large number of pixels.  The intercept is $29.9 \pm 5.9$~\kJysr\ which
compares very well with the $28.3 \pm 0.9$~\kJysr\ found in the ``dark
spot''.

The 17 pixels are not independent because a given star affects several
pixels.  The ratio of the $0.7^\circ$ square beam solid angle to the
pixel solid angle is 4.7:1.  But the effective beam has fuzzy edges where
$p_{ij}$ is close to 0.5, and the appropriate beam solid angle to use is the
noise effective solid angle $\Omega_e = (\sum_i p_{ij})^2/\sum_i p_{ij}^2
= 6$ pixels \citep{Wr85}.
We therefore multiply the statistical uncertainty given above
by a factor of $\sqrt{6} \approx 2.5$,
giving the error bars on the $m < 9$ contribution listed in
Table \ref{tab:decomp}.  But this uncertainty also includes the $\sqrt{N}$ 
fluctuations in the faint stars and any DIRBE instrument noise.

We have run the \citet{WCVWS92} star count model for the ``dark spot'' and
our 2MASS region.  In the 2MASS region the model predicts 785 stars
with $9 < K < 12$~mag and we find 704 stars.  Thus the model count
is $11.5\% \pm 3.8\%$ too high.
The predicted integrated intensity from stars with $9 < K < 12$~mag is
$11.2\% \pm 5.4\%$ too high.
This 11\% difference is well within the 10-15\% estimated uncertainty
in the model \citep{AOWSH98}. 
In the absence of data similar to 2MASS at 3.5 \um, we assume that the
model colors are correct, and 
apply this correction factor to the model values for the
contribution from stars fainter than $9^{th}$ magnitude
in the ``dark spot'' at both 2.2 \um\ and 3.5 \um\ and get
$F = 11.9 \pm 0.6$~\kJysr\ at K and $F = 5.7 \pm 0.3$~\kJysr\ at L.

There is a bias in $F$ introduced by our selection of a ``dark spot''.
Let the total flux which we select on be $T = B+F$, where the bright
star contribution $B$ and the faint star contribution $F$ are independent
random variables.  If $B$ and $F$ are Gaussian, and we select on $T$,
then there is a bias in $F$ of 
$\sigma_F \times (\sigma_F/\sigma_T) \times (T/\sigma_T)$.
We have selected a dark spot that is a fluctuation with probability of
$\sim 10^{-4}$.  The actual fluctuations are not Gaussian, 
and $B$ has infinite variance,
but if we limit $T$ to stars fainter than the 15 Jy cutoff used
by \citet{AOWSH98} a Gaussian approximation is not too bad.
Then $P = 10^{-4}$ is a $-4\sigma$ selection. 
At 2.2 \um, $\sigma_T = 10.8\;\kJysr$ and $\sigma_F = 0.66\;\kJysr$ in
a $2^\circ \times 2^\circ$ box, leading to a bias in $F$ of
$0.66 \times (0.66/10.8) \times (-4) = -0.16\;\kJysr$.
At 3.5 \um, $\sigma_T = 8.8\;\kJysr$ and $\sigma_F = 0.32\;\kJysr$,
leading to a bias in $F$ of $-0.05\;\kJysr$.
We have not corrected for these biases, but in principle our CIRB results
should be increased by $0.16\;\kJysr$ at 2.2 \um\ and $0.05\;\kJysr$ at
3.5 \um.

There is an ISM component at 3.5 $\mu$m that also needs to be subtracted. 
It is 0.00183 times the difference between the zodi-subtracted 100 $\mu$m
intensity, which has a mean of $I_{100} = 1.02\;\MJysr$ in 
our ``dark spot'', and the zero $N_H$
intercept of the $I_{100}$ {\it vs.} $N_H$ correlation, which is
$I_\circ = 0.66\;\MJysr$ \citep{AOWSH98}.
Our maps used for the pixel by pixel bright star fitting 
had 0.00183 times the {\it entire} zodi-subtracted 100 $\mu$m intensity 
subtracted without first removing $I_\circ$.
Furthermore, $I_\circ$ using our zodi model would be $0.4\;\MJysr$ since we
subtracted more zodi.
Since we used $I_{3.5} - 0.00183 I_{100}$ but want
$I_{3.5}-0.00183(I_{100}-I_\circ)$, we need to add a constant
$0.00183 I_\circ = 0.7\;\kJysr$ to our 3.5 \um\ results to complete the ISM
correction.

Noise due to the DIRBE detectors would be included in the pixel to 
pixel scatter, but independent from pixel to pixel.  
Thus the contribution from DIRBE noise
should not be multiplied by the factor of 2.5 discussed above, and we have been
conservative by assuming that the DIRBE noise is zero.  There is no calibration 
error between DIRBE and the zodiacal light models since the models are fit
to data on the DIRBE scale.

Other errors could potentially arise from the zodiacal modeling 
uncertainties and the cross 
calibration uncertainties between the ground based and space based data. 
These uncertainties are dominated by errors in the zodiacal cloud
model, and we estimate them to be 3.8~\kJysr\ at 2.2 \um\ and 
3.3~\kJysr\ at 3.5 \um.
Taking the quadrature sum of 2.5 times the statistical error and our
estimated zodiacal modeling errors gives a net uncertainty of 
\KkJysig~\kJysr\ at 2.2 \um\ and \LkJysig~\kJysr\ at 3.5 \um.

Having determined the stellar contribution we removed it from the DIRBE map 
arriving at a residual signal at K of:
\begin{equation}
I_{2.2 \mu \rm m~} = \KkJy
\end{equation}
or
\begin{equation}
\nu I_{2.2 \mu \rm m~} = \KnW
\end{equation}
and a signal at L of:
\begin{equation}
I_{3.5 \mu \rm m~} = \LkJy
\end{equation}
or
\begin{equation}
\nu I_{3.5 \mu \rm m~} = \LnW
\end{equation}
Table \ref{tab:decomp} shows the total DIRBE intensities and the zodiacal
light, interstellar medium, bright stars, faint stars, and the residual 
signal. 
It should be emphasized that using the \citet{KWFRA98} zodiacal light model 
instead of our model gives a residual intensity that is 5.1~\kJysr\ higher 
at 3.5 $\mu$m and 6.6~\kJysr\ higher at 2.2~$\mu$m.

Using the DIRBE faint source model calibration of 
$F_\circ(L) = 285\;\mbox{Jy}$
instead of the $F_\circ(L) = 263\;\mbox{Jy}$ we measured
would reduce the derived residual intensity at 3.5 \um\ by 0.74~\kJysr.

\section{Discussion}

The technique of zodiacal light subtraction used in this paper reduces
the possibility that this residual signal is an artifact of this model. 
Certainly replacing our zodi model by the \citet{KWFRA98} model would
increase the apparent statistical significance of our result, 
but would leave an unexplained high latitude residual in the 25 \um\ map.
Our mapping of the dark spot also eliminates the possibility that the 
signal is stellar in origin:
by observing and directly subtracting stars down to $9^{th}$ magnitude
we reduce our dependence on the 
\citet{WCVWS92} starcount model by a factor of 4 relative
to the analysis of \citet{AOWSH98} and \citet{HAKDO98}.
Hence we conclude that the residual signal is the CIRB.

Since we have only observed 0.01\% of the sky, we cannot directly verify the
isotropy of the CIRB that we see.  But the evidence we do have for isotropy
at 3.5 \um\ is qualitatively the same kind of evidence 
that \citet{HAKDO98} had for isotropy of
CIRB at 240 \um.  The 240 \um\ map is {\em NOT} isotropic, but
$I_{240} - R(240)(I_{100} - I_\circ)$ is.  Finding the level of this
isotropic signal requires a determination of $I_\circ$ by carefully 
subtracting the galactic signal in a ``dark spot'': the Lockman hole.
DA98 show that $I_{3.5} - 0.312(I_{2.2} - I_\circ)$ is isotropic
except for a low-level large-scale gradient caused by an error in the zodiacal 
light model.  DA98 did not give the magnitude of the zodiacal gradient that
they saw at 3.5 \um, but \citet{WR00} find the slope \vs\ $\csc\beta$
to be 3.1~\nWmmsr\ in the ZSMA maps used by DA98 and 2.3~\nWmmsr\ in the maps
used in this paper.  This 3.5 \um\ slope
is less than the 3.6~\nWmmsr\ slope in the 240 \um\ ISM2 map
reported on line 15 of Table 2 in \citet{HAKDO98}, 
and much less than the 6.9~\nWmmsr\ standard deviation of the 240 \um\ ISM2
slope.
Thus the bolometer noise in the 240 \um\ ISM2 map could mask a larger 
gradient than the gradient seen in the low noise 3.5 \um\ map.
The slope in the 3.5 \um\ map is 21\% of the CIRB, while the reported
240 \um\ ISM2 slope is $26 \pm 51\;\%$ of the CIRB.
Furthermore, the 2 point correlation function of 
the 3.5 \um\ residual \citep{DA99}, as 
a fraction of the CIRB, is smaller than the uncertainty in the
2 point correlation function of the 240 \um\ ISM2 map.
While DA98 obtained a lower limit on $I_\circ$ using galaxy counts at 2.2 \um,
we have determined the value
of $I_\circ$ by carefully subtracting the galactic signal in our dark spot.
Thus the \citet{HAKDO98} determination of the far-infrared CIRB and our
determination of the near-infrared CIRB are logically equivalent.
But one has more confidence in the \citet{HAKDO98} result because
\citet{AOWSH98} were able to determine their $I_\circ$ at 100 \um\ in 3 
areas with consistent results, and because the zodiacal light (and its 
modeling error) are a much smaller fraction of the CIRB at 240 \um\ than 
at 3.5 \um.  

Although the uncertainty for the 2MASS spot is large, we find that 
the value of the residual signal derived from the ``dark spot''
agrees with the 2MASS spot. 
In the future, we will be able to further verify the isotropic nature of 
this signal. Since 2MASS saturates on stars that are quite faint for DIRBE, 
2MASS data can only be used in regions with no bright stars. 
The current 2MASS data release does not cover any of our DIRBE dark spots.
However, future data releases will eventually cover the whole sky, allowing 
a good isotropy check on the CIRB at 2.2 $\mu$m.

Our accepted proposal to survey the darkest spot in the Southern sky
at the South Pole SPIREX/ABU facility in the 3.5 \um\ band during the 
1999 austral winter was not executed due to equipment problems.  
Surveying other dark spots at 3.5 \um\ would be very valuable but also 
very difficult with ground-based telescopes.

Even though our evidence for the isotropy of the near infrared background
is not strong, the uncertainty in our final result is still dominated by the
uncertainties in modeling the zodiacal light.  Our CIRB values at both
2.2 and 3.5 \um\ are 19\% of the zodiacal light at the ecliptic poles,
while the 240 \um\ background found by \citet{HAKDO98} is 3.6 times larger
than the zodiacal light at the ecliptic poles.  Thus the near IR background
is much more sensitive to zodiacal light errors than the far IR background.

\section{Conclusion}

\begin{figure}[ptb]
\plotone{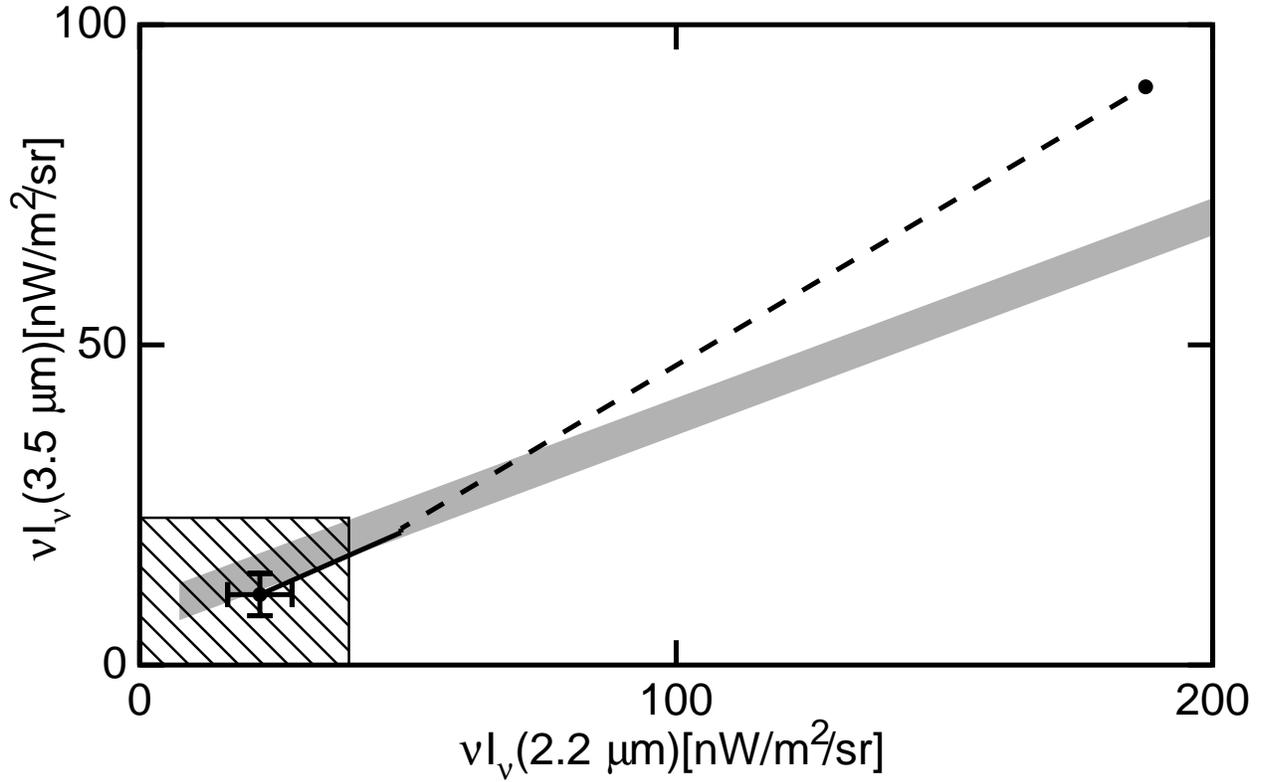}
\caption{
Comparison of our results (point with error bars) and previous limits:
the hatched region shows the \protect\citet{HAKDO98} upper limits, while
the gray bar shows the DA98 correlation.  The point in the upper right shows
the actual average DIRBE intensity in our 17 pixel dark spot, 
with the dashed line showing the zodiacal light
subtraction, and the solid line showing the star light subtraction.
\label{fig:DA98GWC}}
\end{figure}

We compare our values to the values obtained by \citet{HAKDO98}, and  
DA98 in Figure \ref{fig:DA98GWC}. 
\citet{HAKDO98} made no claim for the CIRB since they did not get 
isotropy over a large part of the sky. 
However they did publish values for regions of the sky 
where all foreground contaminations were expected to be lowest. They labeled 
them as high quality (HQ) regions. HQA covered approximately 20\% of the sky 
and HQB covered approximately 2\% of the sky. 
Their HQA value at K was $9.5 \pm 8.8 \;\kJysr$ ($13 \pm 12\;\nWmmsr$) while 
their HQB values was $10.9 \pm 8.8 \; \kJysr$ ($14.9 \pm 12\;\nWmmsr$), 
where the uncertainties are the systematic errors. 
Our value falls within one sigma of both their HQA and HQB values though 
that is due mostly to the fact that their error bars are very large. 

At L band their HQA value was $13 \pm 7 \; \kJysr$ ($11 \pm 6 \;\nWmmsr$) 
and their HQB value was $13.3 \pm 7 \; \kJysr$ ($11.4 \pm 6 \; \nWmmsr$).
Our L band value is also within one sigma though much closer to their 
quoted numbers.

Since DA98 were able to make a much more precise measure than \citet{HAKDO98},
the comparison between our value and theirs is very interesting. 
Inserting our value for $\nu I_\nu(2.2\;\mu\mbox{m})$ 
into the DA98 correlation predicts 
$\nu I_\nu(3.5\;\mu\mbox{m}) = (14.6 \pm 3.5)\; \nWmmsr$ 
which is one standard deviation higher than our result.
Note that DA98 essentially only determined that the high galactic latitude
intensity could not be entirely due to galactic stars because a 
smooth component with a redder K-L color was needed.
In principle, this ``horse of a different color'' could have been
incompletely subtracted zodiacal light, since the K-L color of the zodiacal
light is much redder than typical stars due to the contribution of thermal
emission to the 3.5 $\mu$m intensity.
If we had used the \citet{KWFRA98} zodiacal light model instead of our own,
the inferred 2.2 $\mu$m CIRB would have increased by 6.6~\kJysr
and the 3.5 $\mu$m CIRB would have increased by 5.1~\kJysr.
These changes would have induced a change of
$3\times(5.1/3.5\;-\;0.312\times6.6/2.2) = 1.6\;\nWmmsr$
in the difference between the
DA98 result and our result, in the sense that the difference would go from
$(3.6 \pm 3.5)\; \nWmmsr$ to $(2.0 \pm 3.5)\; \nWmmsr$.
Thus our values are in very good agreement with the DA98 correlation.
Since we use a different zodiacal light model and a different method of star 
subtraction, the fact that we and DA98 have arrived at similar numbers gives
confidence that the residual signal we derive 
at 2.2 \um\ and 3.5 \um\ is very likely the CIRB rather than an unknown
component of Galactic emission.

\acknowledgments

The {\sl COBE} datasets were developed by the NASA Goddard Space Flight
Center under the guidance of the COBE Science Working Group and were
provided by the NSSDC.  This publication makes use of data products from the
Two Micron All Sky Survey, which is a joint project of the University of
Massachusetts and the Infrared Processing and Analysis Center, funded by the
National Aeronautics and Space Administration and the National Science
Foundation.

\appendix

\section{Zodiacal Light Model}
\label{app.zodi}
The zodiacal light model used in this paper is derived from the model described
in \citet{Wr98}.  Three changes have been made:
\begin{enumerate}
\item
The scattering phase 
function coefficients $p_{20}$ and $p_{21}$ are allowed to be different for
each of the three short wavelength bands.  In Table \ref{tab:zodicloud},
I use JKL to distinguish the values for the 1.25, 2.2 and 3.5 $\mu$m bands.
\item
The scaling of the IRAS zodiacal bands, $p_{13}$, is allowed to be different
for the scattered light, denoted $p_{13S}$, than it is for the thermal
emission, denoted $p_{13T}$.
\item
The model was fit to the ``very strong no zodi'' condition of \citet{Wr97}.
\end{enumerate}
The emissivities are given in Table \ref{tab:zodiemiss}
and the IRAS band parameters are given in Table \ref{tab:bandpar}.

\clearpage

\begin{deluxetable}{rlllllll}
\tablecaption{Photometry of Stars in the Dark Spot
with $K < 9$ mag or $L < 9$ mag\label{tab:stars}}
\tablehead{
\colhead{Name} & \colhead{$\alpha_{2000}$} & \colhead{$\delta_{2000}$} &
\colhead{LIRC2} & \colhead{J60} & \colhead{K60} & \colhead{K} & \colhead{L}
}
\startdata
D48  & $13^h02^m53.7^s$ & $ 51^\circ02^\prime21^{\prime\prime}$ &
    7.31 &    6.69 &    6.02 &    6.06 &    5.83 \\
C61  & $12^h56^m31.5^s$ & $ 51^\circ04^\prime49^{\prime\prime}$ &
    6.36 &    7.11 &    6.36 &    6.42 &    6.17 \\
A96  & $13^h03^m09.1^s$ & $ 51^\circ42^\prime28^{\prime\prime}$ &
    6.83 &    7.19 &    6.55 &    6.65 &    6.48 \\
C250 & $12^h52^m21.4^s$ & $ 50^\circ16^\prime56^{\prime\prime}$ &
    7.63 &    7.78 &    7.51 &    7.58 &    7.51 \\
D240 & $12^h59^m31.9^s$ & $ 50^\circ21^\prime47^{\prime\prime}$ &
    8.39 &    8.89 &    8.22 &    8.22 &    7.67 \\
B192 & $12^h54^m45.0^s$ & $ 51^\circ33^\prime50^{\prime\prime}$ &
    8.08 &    8.43 &    7.89 &    7.97 &    7.72 \\
A207 & $12^h58^m17.5^s$ & $ 51^\circ16^\prime28^{\prime\prime}$ &
    8.11 &    8.82 &    8.04 &    8.13 &    7.72 \\
B262 & $12^h52^m47.8^s$ & $ 51^\circ12^\prime09^{\prime\prime}$ &
    8.14 &    8.56 &    8.29 &    8.32 &    7.84 \\
A210 & $12^h58^m17.5^s$ & $ 51^\circ16^\prime28^{\prime\prime}$ &
    9.25 &    8.84 & \nodata &    7.42 &    7.87 \\
D78  & $12^h57^m30.5^s$ & $ 50^\circ49^\prime26^{\prime\prime}$ &
    8.10 &    8.47 &    7.96 &    8.04 &    7.88 \\
D74  & $12^h56^m52.3^s$ & $ 50^\circ48^\prime05^{\prime\prime}$ &
    7.72 &    8.45 &    7.86 &    7.94 &    7.89 \\
A13  & $13^h00^m00.2^s$ & $ 52^\circ08^\prime03^{\prime\prime}$ &
    8.45 &    9.49 &    8.82 &    8.85 &    8.11 \\
C199 & $12^h50^m56.4^s$ & $ 50^\circ29^\prime56^{\prime\prime}$ &
    8.02 &    8.75 &    8.41 &    8.48 &    8.32 \\
A57  & $13^h01^m21.1^s$ & $ 51^\circ51^\prime50^{\prime\prime}$ &
    8.82 &    9.55 &    8.79 &    8.87 &    8.38 \\
A107 & $12^h59^m44.0^s$ & $ 51^\circ42^\prime11^{\prime\prime}$ &
    8.84 &    9.94 & \nodata &    8.66 &    8.42 \\
B257 & $12^h52^m25.8^s$ & $ 51^\circ14^\prime13^{\prime\prime}$ &
    9.02 &    9.41 & \nodata &    9.04 &    8.44 \\
B254 & $12^h51^m50.4^s$ & $ 51^\circ16^\prime29^{\prime\prime}$ &
    8.59 &    8.83 &    8.52 &    8.52 &    8.46 \\
A86  & $13^h02^m56.6^s$ & $ 51^\circ48^\prime14^{\prime\prime}$ &
    8.83 &    9.89 & \nodata &    8.59 &    8.47 \\
B267 & $12^h54^m52.6^s$ & $ 51^\circ13^\prime04^{\prime\prime}$ &
    9.16 &    9.92 & \nodata &    9.74 &    8.50 \\
C261 & $12^h54^m24.7^s$ & $ 50^\circ15^\prime43^{\prime\prime}$ &
    8.80 &    9.32 &    8.75 &    8.81 &    8.72 \\
B50  & $12^h56^m43.5^s$ & $ 52^\circ03^\prime08^{\prime\prime}$ &
    8.62 &    8.92 &    8.62 &    8.62 &    8.78 \\
A175 & $13^h00^m40.4^s$ & $ 51^\circ24^\prime27^{\prime\prime}$ &
    9.00 &    9.23 & \nodata &    8.97 &    8.88 \\
A117 & $12^h57^m52.1^s$ & $ 51^\circ37^\prime56^{\prime\prime}$ &
    9.14 &    9.71 & \nodata &    9.21 &    8.90 \\
C211 & $12^h53^m28.5^s$ & $ 50^\circ28^\prime24^{\prime\prime}$ &
    8.86 &   10.39 & \nodata &    8.82 &    8.91 \\
C49  & $12^h52^m44.3^s$ & $ 51^\circ03^\prime23^{\prime\prime}$ &
    8.87 &   10.37 & \nodata &    9.02 &    8.93 \\
D221 & $13^h02^m37.7^s$ & $ 50^\circ24^\prime44^{\prime\prime}$ &
    8.62 &    9.31 &    8.88 &    8.88 &    9.04 \\
B91  & $12^h51^m47.5^s$ & $ 51^\circ51^\prime15^{\prime\prime}$ &
    8.28 &    8.44 &    8.16 &    8.19 &    9.07 \\
D60  & $13^h01^m03.8^s$ & $ 50^\circ59^\prime28^{\prime\prime}$ &
    8.90 &    9.13 & \nodata &    8.79 &    9.51 \\
\enddata
\end{deluxetable}

\begin{deluxetable}{lrr}
\tablecaption{Decomposition of the DIRBE Intensity \label{tab:decomp}}
\tablehead{
\colhead{Component} &
\colhead{$2.2\;\mu$m} &
\colhead{$3.5\;\mu$m} \\
\colhead{} & \colhead{kJy sr$^{-1}$} & \colhead{kJy sr$^{-1}$}
}
\startdata
Total              &   $137.5 \pm 0.3$ & $105.3 \pm 0.3$ \\
Zodi               &   $101.8 \pm 3.8$ &  $80.4 \pm 3.3$ \\
ISM                & \nodata           &   $1.1 \pm 0.2$ \\
Stars, $m < 9$ mag &     $7.4 \pm 2.2$ &   $5.3 \pm 1.8$ \\
Stars, $m > 9$ mag &    $11.9 \pm 0.6$ &   $5.7 \pm 0.3$ \\
CIRB               & $\KkJyval \pm \KkJysig$ & $\LkJyval \pm \LkJysig$ \\
\enddata
\end{deluxetable}

\begin{deluxetable}{ccl}
\tablecaption{Diffuse Cloud Parameters \label{tab:zodicloud}}
\tablehead{
\colhead{Parameter} &
\colhead{Value} & 
\colhead{Description} 
}
\startdata
$p_1$ &        1.2346 & radial density exponent \\
$p_2$ &        0.4246 & radial temperature exponent \\
$p_3$ &        3.5785 & vertical ``scale height'' \\
$p_4$ &        0.9450 & vertical density exponent \\
$p_5$ &       -1.3559 & $\ln(\sin i)$ at break \\
$p_6$ &        0.3838 & $10\times$ cloud pole $x$ component \\
$p_7$ &       -0.0758 & $10\times$ cloud pole $y$ component \\
$p_8$ &       -0.0195 & $10\times$ cloud offset $x$ component \\
$p_9$ &       -0.0471 & $10\times$ cloud offset $y$ component  \\
$p_{10}$ &     0.6098 & $10\times$ density contrast of Dermott ring\\
$p_{11}$ &     5.5405 & $\ln(T_B)$, band temperature at $R=1$ \\
$p_{12S}$ &    0.5722 & scattering scale factor for bands \\
$p_{12T}$ &    1.3851 & thermal scale factor for bands \\
$p_{13}$ &     0.3161 & the ``dimple'' in Dermott's ring \\
$p_{14}$ &     7.8852 & vertical scale for Dermott's ring \\
$p_{15}$ &    -0.0226 & spherical term in vertical density \\
$p_{16}$ &     0.0289 & $\sin^2 i$ term in vertical density \\
$p_{17}$ &    -0.0262 & additional density at $|\sin i| \approx 0.5$ \\
$p_{18}$ &    -0.1977 & additional density at $|\sin i| \approx 0.25$ \\
$p_{19}$ &    -0.0294 & additional density at $|\sin i| \approx 0.17$ \\
$p_{20J}$ &   -0.2616 & phase function linear coefficient at 1.25 $\mu$m \\
$p_{20K}$ &   -0.2865 & phase function linear coefficient at 2.2 $\mu$m \\
$p_{20L}$ &   -0.2634 & phase function linear coefficient at 3.5 $\mu$m \\
$p_{21J}$ &    0.6761 & phase function quadratic coefficient at 1.25 $\mu$m \\
$p_{21K}$ &    0.5318 & phase function quadratic coefficient at 2.2 $\mu$m \\
$p_{21L}$ &    0.3988 & phase function quadratic coefficient at 3.5 $\mu$m \\
\enddata
\end{deluxetable}

\begin{deluxetable}{ccl}
\tablecaption{Scattering and Emission Efficiencies \label{tab:zodiemiss}}
\tablehead{
\colhead{Parameter} &
\colhead{Value} & 
\colhead{Description} 
}
\startdata
$p_{22}$ &     0.6647 & $\sigma_1$ scaling \\
$p_{23}$ &     0.7762 & $\sigma_2$ scaling \\
$p_{24}$ &     0.7499 & $\sigma_3$ scaling \\
$p_{25}$ &     1.8247 & $\kappa_{ 3}$ scaling \\
$p_{26}$ &     1.3150 & $\kappa_{ 4}$ scaling \\
$p_{27}$ &     1.1088 & $\kappa_{ 5}$ scaling \\
$p_{28}$ &     1.2525 & $\kappa_{ 6}$ scaling \\
$p_{29}$ &     0.8657 & $\kappa_{ 7}$ scaling \\
$p_{30}$ &     0.7547 & $\kappa_{ 8}$ scaling \\
$p_{31}$ &     0.8147 & $\kappa_{ 9}$ scaling \\
$p_{32}$ &     0.6130 & $\kappa_{10}$ scaling \\
\enddata
\end{deluxetable}

\begin{deluxetable}{ccl}
\tablecaption{IRAS Band Parameters \label{tab:bandpar}}
\tablehead{
\colhead{Parameter} &
\colhead{Value} & 
\colhead{Description} 
}
\startdata
$q_1$ &   1.7058  & $10\times(\sin i)_{max}$ for Band 1 \\
$q_2$ &   0.1963  & Band 1 normalization \\
$q_3$ &   0.2722  & $10\times(\sin i)_{max}$ for Band 2 \\
$q_4$ &   0.2515  & Band 2 normalization \\
$q_5$ &   0.1167  & $10\times$ band pole $x$ component \\
$q_6$ &  -0.2691  & $10\times$ band pole $y$ component \\
$q_7$ &  -0.9321  & $10\times$ band offset $x$ component \\
$q_8$ &   1.9164  & $10\times$ band offset $y$ component \\
$R_1$ & 3.14 & Outer radius for Band 1 \\
$R_2$ & 3.02 & Outer radius for Band 2 \\
\enddata
\end{deluxetable}

\end{document}